# Wavelets as a variational basis of the $XY$ model

C. Best,[a] [*]   A. Schäfer[a], and W. Greiner[a]

[a]Institut für Theoretische Physik
Johann Wolfgang Goethe-Universität
60054 Frankfurt am Main, Germany

We use Daubechies' orthonormal compact wavelets as a variational basis for the $XY$ model in two and three dimensions. Assuming that the fluctuations of the wavelet coefficients are Gaussian and uncorrelated, minimization of the free energy yields the fluctuation strength of wavelet coefficients at different scales, from which observables can be computed. This model is able to describe the low-temperature phase and makes a prediction about the phase transition temperature.

## 1. Introduction

The success of wavelets in analyzing complex signals has prompted speculation about their possible applications to field theories and other lattice systems. We report here on the first part of an investigation into possible wavelet tools that can be applied to lattice field theories. The wavelet representation of a lattice field is analyzed by a variational principle to obtain an approximate description of a phase transition. Wavelets represent nonlocal fluctuations that eventually induce the phase transition.

### 1.1. The wavelet transform

A wavelet [1,2] $\psi \in \mathbf{L}^2(\mathbf{R})$ is a function whose binary dilatations and dyadic translations generate a Riesz basis of $\mathbf{L}^2(\mathbf{R})$ so that any function $f \in \mathbf{L}^2(\mathbf{R})$ can be expanded into a wavelet series

$$f(x) = \sum_{n=-\infty}^{\infty} \sum_{x' \in \mathcal{L}^n} \hat{f}^{(n)}(x') \, \psi^{(n)}(x')(x) \qquad (1)$$

with basis functions $\psi^{(n)}(x') \in \mathbf{L}^2(\mathbf{R})$,

$$\psi^{(n)}(x')(x) = 2^{-n/2} \psi\left(2^{-n}(x-x')\right) \quad, \qquad (2)$$

labeled by scale $n \in \mathbf{Z}$ and position $x \in \mathcal{L}^n$ on the $n$-th sublattice $\mathcal{L}^n = 2^n \mathbf{Z}$.

Compact orthonormal wavelets as constructed by Daubechies [3] are not given analytically but by a powerful numerical procedure [4,5]. The algorithm acts on vectors (i.e. functions discretized on a lattice), seperating the information contained in a vector into a smoothed vector on a coarser lattice and a complementary detail vector. Repeated application of the algorithm strips the initial vector scale by scale of its detail information (= wavelet coefficients). Being a linear orthogonal transformation, it defines an orthonormal basis consisting of wavelets (associated to the detail information) and scaling functions (associated to the smoothed information). The discrete wavelets defined in this way approach the continuum wavelet.

Wavelets in more than one dimension can be constructed by direct products. They then carry an additional label $t = 1, \ldots, 2^D - 1$ that specifies their composition.

## 2. The model

Our model is the $XY$ spin model defined by the Hamiltonian

$$\mathcal{H} = \sum_{x \in \mathcal{L}} \sum_{\mu=1}^{D} \{1 - \cos(a(x+e_\mu) - a(x))\} \qquad (3)$$

on a lattice $\mathcal{L} = \mathbf{Z}^D$ in $D$ dimensions. The field $-\pi \leq a(x) < \pi$ gives the angle of the spin at site $x$, and $e_\mu$ the unit vector in direction $\mu$. In two dimension, this model undergoes a Kosterlitz-Thouless phase transition between a disordered high-temperature phase and a spin-wave dominated low-temperature phase.

---

[*]email: cbest@th.physik.uni-frankfurt.de

## 2.1. Variational principle

We employ a variational principle to obtain an approximate description of the probability distribution in the partition sum. To this end, the field $a(x)$ is expanded in a wavelet basis. We choose a Gaussian trial probability distribution characterized by the correlation function

$$\langle \hat{a}_t^{(n_1)}(x_1) \hat{a}_t^{(n_2)}(x_2) \rangle = \delta_{n_1,n_2} \delta_{x_1,x_2} \mathcal{A}_t^{(n_1)} \quad , \quad (4)$$

i.e., fluctuations of the *wavelet coefficients* are assumed to be uncorrelated. The values of the variational parameters $\mathcal{A}_t^{(n)}$, describing the strength of fluctuations at scale $n$, are obtained by calculating the internal energy $U$ (expectation value of $\mathcal{H}$ in this ensemble) and minimizing the free energy

$$F = U - \frac{S}{\beta} \quad (5)$$

Due to the Gaussian ansatz, both the entropy $S$ and the internal energy $U$ can be easily calculated. However, the field $a(x)$ can now assume any number although the Gaussian ansatz does not reflect the periodicity of the Hamiltonian. This will limit the applicability of the variational method to the low-temperature phase where periodicity is not important.

## 2.2. Calculation of the internal energy

We denote the wavelet transform of the field $a(x)$ by $\hat{a}^{(n)}(x')$, $x' \in \mathcal{L}^n$, defined on the sublattices $\mathcal{L}^{(n)} = (2^n \mathbf{Z})^D$. To obtain the Hamiltonian in terms of the wavelet-transformed field, we must express the difference of two adjacent spin angles in wavelet coefficients. This is done using the characteristic function of the link,

$$L(x,\mu)(y) = \delta_{y,x+e_\mu} - \delta_{y,x} \quad , \quad (6)$$

and the scalar product in the space of lattice fields. In wavelet space, the difference is a scalar product of the wavelet transform $\hat{L}_t^{(n)}(x,\mu)(x')$ of the link with the wavelet coefficients of the field. Thus the simple difference of two adjacent lattice sites is replaced by a more complicated, albeit still linear, expression, the cosine of which can be evaluated in a Gaussian ensemble. One finds for the internal energy

$$U = \sum_{x,\mu} \left[ 1 - \exp\left( -\sum_{n,t} |L_t^{(n)}(x,\mu)|^2 \mathcal{A}_t^{(n)} \right) \right] \quad (7)$$

The quantity

$$|L_t^{(n)}(x,\mu)|^2 = \sum_t \sum_{x' \in \mathcal{L}^n} \left( L_t^{(n)}(x,\mu)(x') \right)^2 \quad (8)$$

characterizes the power spectrum of the link $(x,\mu)$, i.e., how much of the original square norm ("power") of the link's characteristic function is on scale $n$. It can be computed numerically and determines how much fluctuations at this scale contribute to the internal energy at this link.

## 3. Results

### 3.1. Free fields

The same procedure as above can be performed for a Gaussian model (free field theory), corresponding to expanding the cosine and retaining only the first term. In this case, the minimization can be performed analytically yielding

$$\mathcal{A}_t^{(n)} = \frac{1}{2\beta} \frac{N_n}{\sum_x \sum_{\mu=1}^{D} |\hat{L}_t^{(n)}(x,\mu)|^2} \quad (9)$$

($N_n = \#$ of sites on $\mathcal{L}^n$). Using the scaling properties of wavelets, these obey a simple scaling law from which the correlation function $C(d)$ can be calculated:

$$\mathcal{A}^{(n)} \sim \frac{4^n}{2\beta} \quad \Longrightarrow \quad C(d) \sim d^{-(D-2)} \quad , \quad (10)$$

exhibiting the expected scaling behavior for a free field theory. Note that this critical correlator results from local (noncritical) correlations in wavelet space. This reflects a general hope that wavelets may provide an approximately diagonal basis for some systems.

### 3.2. *XY* model

We have minimized numerically the free energy with respect to the fluctuation strengths $\mathcal{A}_t^{(n)}$. At low temperatures, the model approaches the Gaussian model whose results serve as a starting and reference point. Spins are approximately aligned, and spin wave fluctuations rarely make a single spin turn by a large angle. The periodicity of the Hamiltonian is thus not felt by the model.

At higher temperature, the probability distribution expands and starts to feel the periodicity. At some point, fluctuations become so strong that full turns of the spins are occurring regularly, making a description by a localized Gaussian unsuitable. This is reflected by the fact that the free energy has no absolute minimum but only a relative one which vanishes at a critical temperature.

We interpret this behavior as a phase transition by ergodicity: If we imagine each spin attached to a rubber band that accumulates the total rotation angle, fluctuations in the low-temperature phase do not reach over the crest of the cosine potential, and spins will only rarely accumulate large rotations. The system is thus localized in a small region of phase space. In the high-temperature phase, spins frequently turn and will eventually reach any amount of rotation, spreading out the system periodically over the whole phase space.

Wavelets do not describe the absolute direction of the spins and thus keep global $O(1)$ symmetry. They make it possible to obtain nonlocal correlations using only a small number of variational parameters. However, the $XY$ model is dominated by spin waves which are quite well described by plane waves, and the ability of wavelets to describe complex irregular systems does not come into play in this problem.

We find a critical inverse temperature of $\beta_{\text{crit}} \approx 0.69$ for two dimensions on a $32 \times 32$-lattice and $\beta_{\text{crit}} \approx 0.45\ldots0.49$ in three dimensions. (The accuracy is limited by the shallowness of the relative minimum.) Low-temperature analysis of vortices predicts in two dimensions $\beta_{\text{crit}} = 2/\pi \approx 0.64$.

Fig. 1 shows the internal energy calculated by a Monte Carlo simulation and by the variational approach.

## 4. Conclusions

We have shown how wavelets can be used to describe fluctuations in a statistical field theory. Using a rather small number of variational parameters, an approximate description of fluctuations in the $XY$ model was obtained which is sufficiently accurate to predict a phase transition.

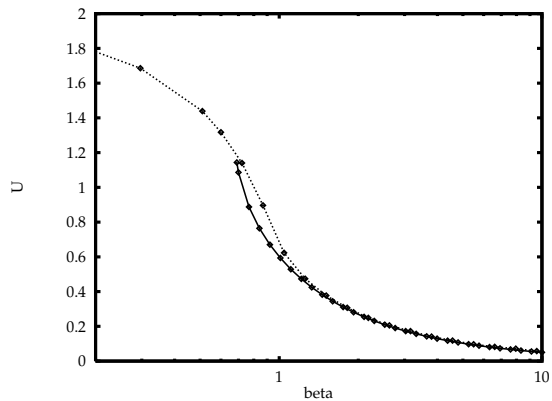

Figure 1. Internal energy per site in the $XY$ model vs. inverse temperature from variational wavelets (solid line) and from Monte Carlo calculation (dotted line).

The resulting phase transition temperatures are in good agreement with their expected values.

Clearly, the approach is limited by the variational principle used and the choice to expand in angle variables instead of spins. Further investigations are in progress to extend the treatment to $SO(N)$ symmetric Landau-Ginzburg theories. As these theories are more easily accessible to position-space renormalization group transformations, we are trying to represent such transformations in wavelet space, effectively integrating out wavelet coefficients at the lowest scale.